# A Security Analysis Of IoT Encryption: Side-Channel Cube Attack On Simeck32/64

Alya Geogiana Buja[1,2], Shekh Faisal Abdul-Latip[1] and Rabiah Ahmad[1]

[1] INSFORNET, Center for Advanced Computing Technology, Universiti Teknikal Malaysia Melaka, Hang Tuah Jaya, Durian Tunggal, 76100 Melaka, Malaysia

[2] Universiti Teknologi MARA, Shah Alam, 40450 Selangor, Malaysia

## ABSTRACT

*Simeck, a lightweight block cipher has been proposed to be one of the encryption that can be employed in the Internet of Things (IoT) applications. Therefore, this paper presents the security of the Simeck32/64 block cipher against side-channel cube attack. We exhibit our attack against Simeck32/64 using the Hamming weight leakage assumption to extract linearly independent equations in key bits. We have been able to find 32 linearly independent equations in 32 key variables by only considering the second bit from the LSB of the Hamming weight leakage of the internal state on the fourth round of the cipher. This enables our attack to improve previous attacks on Simeck32/64 within side-channel attack model with better time and data complexity of $2^{35}$ and $2^{11.29}$ respectively.*

## 1. INTRODUCTION

Internet of Things (IoT) is another improvement of information technology. Internet of things connects not only the computer devices but also the living things like plants, people and animals [26]. The number of connected devices is increasing rapidly that can lead to both opportunity and threats. Therefore, the security of IoT has become a crucial concern among the researcher over the world. Haroon et al. [27] have addressed the technical challenges of IoT. Due to constraints such as connection setup, energy, power, and storage in IoT connected devices; a lightweight encryption is required to secure the IoT communication (see Figure 1). Therefore, this paper presents the approach and security analysis of a to-be IoT encryption.

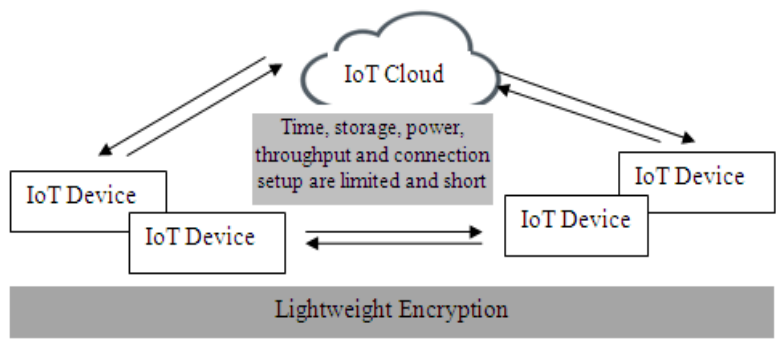

Figure 1. Security landscape of IoT

Simeck32/64 [1] is a lightweight block cipher that was designed based on the combination of good design components from SIMON and SPECK block ciphers [2]. It is an Addition-Rotation-XOR lightweight block cipher. However, Simeck block cipher does not have the modular operation. Nalla et al. have analyzed Simeck32/64 by using fault attack and 16 bits of the last round key have been recovered successfully [3]. As presented in Table 1 in [1], the hardware implementations of the Simeck block cipher family are even smaller than our implementations of SIMON in terms of area and power consumption. The design rationale of the Simeck block cipher suits the requirement of IoT embedded devices such as in RFID tags. In a fault attack, it is assumed that there is a fault occurred during the encryption. However, in the side-channel cube attack, it is assumed that there is a leakage occurred in the cryptosystem. Therefore, in this paper, by using the proposed framework with Hamming-weight leakage bit after four rounds of encryption, this attack has been able to decrease the time complexity of the previous results to $2^{35}$ computations.

Cube attack is a generic type of algebraic attack introduced by Dinur and Shamir at EUROCRYPT in 2009 [4]. Most of the cryptosystems can be represented by a system of multivariate polynomial equations over a finite field, GF(2) .To apply a cube attack, the adversary requires a black-box access to a target cryptosystem and it is assumed that the adversary has an access to a bit of information from the cryptosystem. The obtained information from the cryptosystem enables the adversary to achieve the goal of cube attack which is the adversary can derive low-degree equations that can be exploited for constructing the distinguishers [5] and key recovery attack [4]. When using the original cube attack [4], the adversary tries to derive independent linear equations over secret variables of the cryptosystem. The system of several linear equations can be easily solved to recover the value of the secret variables by using the Gaussian Elimination. Several lightweight block ciphers have been analysed susceptible to cube attacks such as KATAN [6], NOEKEON [7] and PRESENT [8][9]. The cube attack presented in this chapter is motivated by the observation of SIMON (Beaulieu et al., 2013) and KATAN [10] family of block ciphers against algebraic cube attack [6].

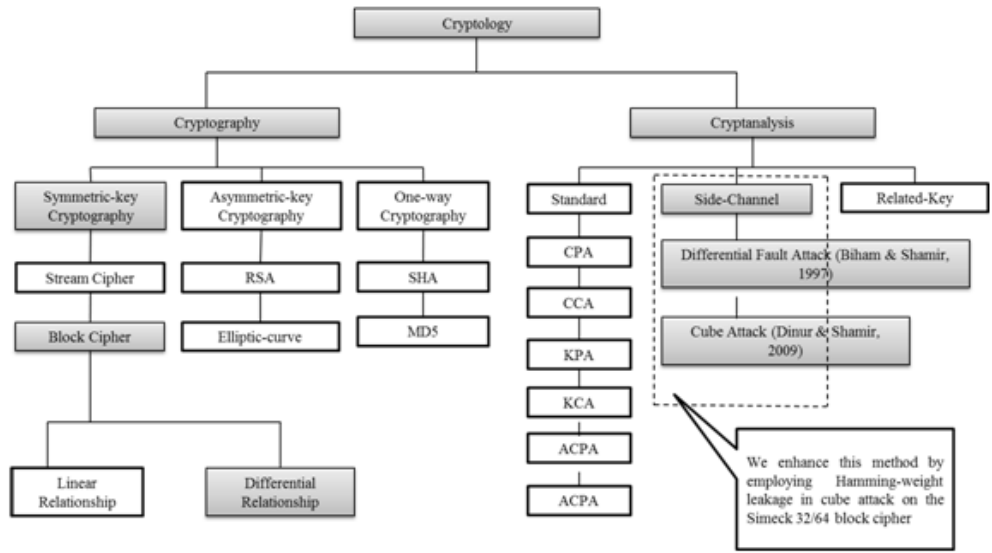

Figure 2. View of Security Analysis

**Our contribution.** In this paper, we propose the employing of Hamming-weight leakage model in cube attack on Simeck32/64. By using this attack model, more sub key bits have been able to be recovered; therefore, our method has been able to improve the previous time complexity from $2^{48}$ to $2^{35}$ in the side-channel model of attack (see Figure 2 for the focus of this analysis).

## 2. RELATED WORKS

Simeck is a block cipher designed based on SIMON [1]. SIMON [2] and KATAN [10] have been analyzed using dynamic cube attack [6] in standard model of attack. In the attack, the researchers use a cube tester which is positioned at the middle of the cipher. The cube tester is extended in two directions over the maximum possible upper and lower rounds provided that some of sub key bits are successfully guessed. The automated algorithm in dynamic cube attack can be realized and the results show that the method can break 118 and 155 out of 254 rounds of KATAN32 in the non-full codebook and full-codebook attack scenarios, respectively. For SIMON32/64, they are able to break 17 and 22 out of 32 rounds, in the same scenarios. In addition, in 2013, [11] and [12] have analyzed SIMON with differential and linear cryptanalysis.

Dinur and Shamir have proposed a side-channel attack model [4]. In the side-channel model, the adversary is assumed to have access to only a bit of information (a leakage bit) about the internal state of the block cipher after each round. The one bit of information can be a single bit of the internal state or a Hamming-weight bit from the internal state. Dinur and Shamir [4] have shown that by using cube attack in the single-bit-leakage side-channel model can recover the secret key of the AES [13] and [14] and SERPENT block ciphers much easier than the previously known side-channel attacks. Yang et al. have investigated PRESENT block cipher using side-channel cube attack [8]. A year later, the side-channel cube attack has been applied to NOEKEON [9] and PRESENT [7][8][15]. For the NOEKEON block cipher, the complexity of the previous attack is reduced to $2^{68}$ computations in single-bit leakage model. 60 linearly independent equations over 99 key variables have been extracted successfully. Meanwhile, for the PRESENT block cipher, Abdul-Latip et al. [7] have been able to reduce the attack complexity to $2^{16}$ computations with $2^{18}$ chosen plaintexts for PRESENT-128 and $2^{64}$ with $2^{18}$ chosen plaintexts.

Simeck is a newly introduced cipher in 2015 [1]. Since proposed, [22] [16] [17] and [18] have analyzed Simeck by using linear, differential and impossible differential cryptanalysis too. Later, Zhang et al. [19] and [20] have analyzed Simeck family of block ciphers by using integral cryptanalysis method. The attacks have been able to obtain 12/14/16-round theoretical integral distinguishers on Simeck32/48/64 and some 15-round experimental integral distinguishers on Simeck32. In addition, Xiang et. al. [21] also analyzed Simeck by using integral cryptanalysis and 15, 18 and 21-round distinguishers are found for Simeck32/64, Simeck48/96 and Simeck64/128 respectively. Nalla et al. [3] have investigated Simeck using fault analysis. Firstly, the researchers have applied random bit-flip fault attack and n-bits of the last round key of Simeck have been recovered by using about n/2 faults. Secondly, a random byte fault attack is used and the attack was able to recover the n-bit of last round key of Simeck using about n/6.5 faults.

Therefore, one of the contributions of this paper is providing the security analysis of the Simeck32/64 block cipher against side-channel cube attack. Table 1 shows some results of security analysis on Simeck32/64 against several cryptanalytic methods in side-channel and the standard model of attack.

Table 1. Some results of security analysis on Simeck32/64.

| Attack model | Type of attack | Attack complexity | # of attacked rounds | Reference |
|---|---|---|---|---|
| Side-channel | Cube attack with Hamming-weight leakage model | $2^{35}$ | Full round | This paper |
| | Fault analysis with single-bit fault model | $2^{48}$ | Full round | [3] |
| | Fault analysis with random-byte fault model | $2^{48}$ | Full round | [3] |
| Standard | Linear cryptanalysis | $2^{31}$ | 13 | [22] |
| | Differential cryptanalysis | $2^{31}$ | 19 | [18] |
| | Impossible differential | $2^{32}$ | 20 | [8] |
| | Zero-correlation linear cryptanalysis | $2^{32}$ | 20 | [16] |
| | Differential cryptanalysis with dynamic key guessing | $2^{32}$ | 22 | [17] |
| | Linear-hull cryptanalysis | $2^{31.91}$ | 23 | [20] |

## 3. DESCRIPTION OF SIMECK32/64

Simeck32/64 is a variant of Simeck family of block ciphers (cf. [1] for detailed the explanation about the Simeck block cipher). The design of Simeck32/64 is based on Feistel and Addition-Rotation-XOR (ARX) network which adopted some good components of two NSA ciphers, SIMON and SPECK. The block cipher accepts 32 bits plaintext as the input, 64 bits secret key and 32 rounds for a complete encryption process. In each round, 16-bit sub key is required for the encryption.

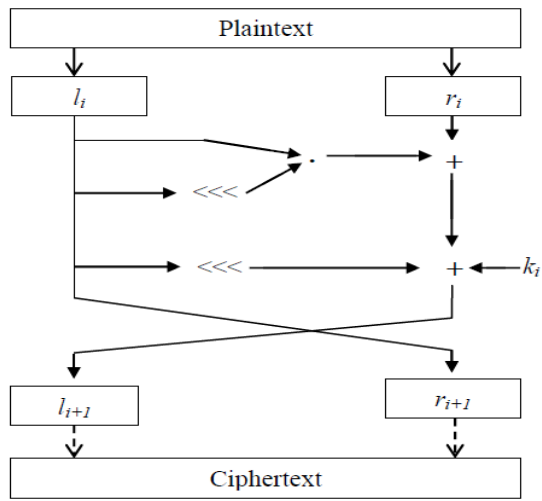

Figure 3. Structure of Simeck for round "i"

Based on the results obtained in a study [18] that has been conducted on the SIMON block cipher, Kolbl and Roy [18] explicitly state the Simeck32/64 block cipher requires 8 rounds to achieve full diffusion; means that each bit at the input affects all bits of the output. The Simeck key schedule is designed using SPECK round function. There are six operations in a round of Simeck round function; ANDing, Rotate left 5 bits, Rotate left 1 bit, three XORing operations (the intermediate state with round key) and finally the swapping process which is immediately takes place before the next round. Figure 3 shows the structure of Simeck at round "i".

# 4. CUBE ATTACK

Cube attack is a generic type of algebraic attack (higher order differential) that was proposed by Dinur and Shamir at EUROCRYPT 2009 [4]. The aim of the cube attack is to recover the secret key in a cryptosystem by extracting and solving linearly independent algebra equations [4]. For a well designed cipher, an algebraic representation over GF(2) is of degreed the cube attack will require about 2d computations. [5] have analyzed the Trivium [10] stream cipher using cube attack. In the cube attack, if the degree of the master polynomial is relatively low, then it is possible for the adversary to analysze the cipher faster than by the exhaustive search (brute force attack).

Table 2. The summation of the master polynomial, p.

| $x_1$ | $x_2$ | $x_3$ | Derived Polynomials |
|---|---|---|---|
| 0 | 0 | 0 | $x_4 x_5$ |
| 0 | 0 | 1 | $x_4 + 1 + x_4 x_5$ |
| 0 | 1 | 0 | $x_4 x_5$ |
| 0 | 1 | 1 | $x_3 x_4 + x_3 + x_4 x_5 + x_2 x_3$ |
| 1 | 0 | 0 | $x_4 x_5$ |
| 1 | 0 | 1 | $x_3 x_4 + x_3 + x_4 x_5 + x_1 x_3$ |
| 1 | 1 | 0 | $x_4 x_5$ |
| 1 | 1 | 1 | $x_4 + x_4 x_5 + 1$ |
| $\sum$ | | | $x_4 + x_1 x_3 + x_2 x_3 + 1$ |

The cube attack has a weakness in a case when the degree of the polynomial is too high because the degree that representing a ciphertext bit grows exponentially with the number of rounds in the cipher. Therefore, for a block cipher, the application of cube attack usually becomes ineffective after a few rounds if the adversary executes in a standard attack model. To get a better result using cube attack, the adversary should derive enough number of linearly independent equations which are solvable by Gaussian elimination. However, when considering the practical implementations of a block cipher, in an embedded system (limited in terms of resources and power) such as smart cards, the cube attack will become a stronger attack if implemented in side-channel attack model (cf. next section for more details about side-channel cube attack).

In the real attack, the adversary is able to access to some (leakage) information from the internal state of the cipher. This information can be the timing, electrical power consumption, electromagnetic radiation and probing. Using cube attack, the main observation is, the summation of p (p is the multivariate master polynomial p(v1,……vn, k1……..kn) which representing the output bits of an encryption algorithm over GF(2)) over tI (a monomial term that containing multiplication of all variable in I; $I \subseteq \{1, ..., \ell\}$ where $\ell$ = m (length of plaintext variable) + n (length of key variables)) by assigning all the possible combinations of '0' and '1' to all variable in I and the other value is fixed. By summing the polynomial p over all possible value of variable in tI requires the polynomial to be summed over an even number of times. Consider the following summation (as presented in Table 2) of the given master polynomial p(x1, x2, x3, x4, x5) = x1 x2 x3 x4 + x3 x4 + x3 + x4 x5 + x1 x3 + x2 x3 by choosing I = {1, 2, 3}.

To execute cube attack, firstly, ignore the distinction between the secret and public variables' notations by denoting all variables by xi , • • • , xℓ, where ℓ = m + n. Let $I \subseteq \{1, ..., \ell\}$ be a subset of the variable indexes, and tI denote a monomial term containing multiplication of all the xis with i ∈ I . For cube attack, if the master polynomial p is factorized by the monomial tI, then the resulted equation is shown in Equation 1.

$$p(x1, …, x\ell \,) = tI \, \bullet pS(I) + q(x1, …, x\ell \,) \tag{1}$$

where pS(I), is called the superpoly of tI in p, which does not have any common variable with tI , and each monomial term tI in the remainder polynomial, denoted by q will miss at least one variable from tI. Meanwhile, tI is called a "maxterm" if its superpoly in p is linear polynomial which is not a constant (for instance; degree of pS(I) is equal to1). For cube attack, the main observation is that, if the summation of polynomial p over tI by assigning all the possible combinations of 0 or 1 values to the xi with i ∈ I and fixing the value of all the remaining xi  with i not in I , the resultant polynomial equals to pS(I) of mod 2. For instance, given a master polynomial p is as follows;

$$p(x1, x2, x3, x4, x5, x6, x7) = x1 \, x2 \, x3 + x1 \, x2 \, x3 \, x4 + x2 \, x4 \, x6 + x1 \, x2 \, x3 \, x5 \, x7$$

Then, choose a subset I = {1, 2, 3} where tI = x1 x2 x3 and the resulted tweakable polynomial (as in Equation 1) will be

$$p(x1, x2, x3, x4, x5, x6, x7) = x1 \, x2 \, x3 \, (1 + x4) + (x2 \, x4 \, x6 + x5 \, x7)$$

**Theorem 1** *(The Main Observation [4][7]). Given a master polynomial p over GF(2) with ℓ variables, and any index subset I ⊆ {1, • • •, ℓ}, then  pI = pS(I ).*

Given an access to a cryptographic function with public and secret variables, from this observation, the adversary is able to recover the value of the secret variables (ki) in two phases, namely preprocessing and online phase. In the preprocessing phase, firstly, the adversary has to find sufficient number of maxterms, tI, such that each tI consists of a subset of public variables v1, • • •, vm . To find the maxterms, the adversary has to conduct a linearity test such as BLR [23] on pS(I) over the secret variables ki ∈ {k1, • • •, kn} and the value of the public variables which is not in tI are fixed (to 0 or 1). To conduct this test, the adversary has to choose a sufficient number of vectors x and y ∈ {0, 1}n independently and uniformly at random representing samples of n-bit key. Then, for each pair of vectors x and y, the adversary has to sum the polynomial p over tI to verify whether or not each one of the vectors satisfies Equation 2. If all the vectors x and y satisfy Equation 2, with high probability pS(I) is linear over the secret key variables. The next step is to derive linearly independent equations in the secret variables ki from pS(I). By having and solving these linearly independent equations, the adversary can obtain the values of the cipher key.

$$pS(I) \, [0] + pS(I)[x] + pS(I) \, [y] = pS(I)[x + y] \tag{2}$$

The preprocessing is completed if the adversary has gather sufficient number of linearly independent equations in key variables. In the online phase, the adversary is required find the value of the right-hand side of each linear equation by summing the black box polynomial, p over the same set of maxterms tIs which have been obtained in the preprocessing phase. Finally, the adversary can recover the value of cipher key by solving the linear equations using the Gaussian Elimination.

## 5. A Review on Side-channel Cube Attack

Side-channel cube attack [24] is a hybrid attack that tries to exploit the implementation of full rounds cryptosystem by manipulating the intermediate round information leakage and extend the attack until the last round of the side-channel attack. If the cipher is not properly implemented (especially if the cipher is implemented in an embedded system), by using side-channel cube attack, the adversary is able to access a bit of information of the cipher. Hence, the side-channel cube attack can be a real threat to many block ciphers, such as KATAN [25], NOEKEON [9] and PRESENT [8][7]. In the pre-processing phase of side-channel cube attack, there are two main tasks; firstly, the adversary has to determine the efficient round to launch the attack and secondly, the adversary has to find the maxterm equations. To determine the efficient round, the adversary has to determine the number of key variables in master polynomial after each round (cf. [9] for further description on how to determine the number of key variables). It is recommended to choose round where the cipher achieves full diffusion. If the full diffusion occurred in the cipher, more secret variables can be obtained. In the side-channel attack model, the attack can be launched at any round of the cipher. Next, in the online phase, the adversary has to recover the value of key variables.

## 6. Side-channel Cube Attack on Simeck32/64

This section explains the side-channel cube attack against Simeck32/64 using the Hamming-weight leakage model. In order to apply the Hamming-weight leakage side-channel cube attack on the Simeck32/64 block cipher, firstly, the adversary needs to find out the round in which the cipher begins achieving complete diffusion. This enables the adversary to find most of the key variables from low degree master polynomials in early rounds of the encryption process. Based on Yang et al. [8], the Simeck32/64 block cipher achieves full diffusion after rounds and the Simeck block cipher is claimed susceptible to any algebraic attack after 5 rounds of encryption. However, the degree of master polynomials after 8 rounds is very high which is not suitable for our side-channel attack model. In addition, as mentioned in the Simeck proposal [1], the degree of the polynomial of the Simeck block cipher (any variant) after 5 rounds is 13. Therefore, in this study, the attack is executed by considering the Hamming-weight leakage after 4 rounds of encryption to extract sufficiently many maxterms.

As in Definition 1, in this study, the value of the Hamming-weight of the internal state after 4 rounds is represented 8-bit binary form by considering all bits starting from LSB position towards the MSB position. In this attack, in order to search for sufficient maxterms, two cube sizes are applied (6 and 8) and the second bit of the Hamming-weight is chosen.

**Definition 1** [7] Let B represents the internal state of the cipher and $B = b_{\beta-1} \ldots b_0$ be the binary string of length $\beta$ bits. The Hamming-weight of B is the number of bits with value 1 in the binary representation of B, which can be computed as $HW(B) = \sum_{\beta=0}^{j-1} b_j$ and has a value between 0 and $\beta$. The LSB of $HW(B)$ is the XOR of all bits from B and the MSB of $HW(B)$ is the AND of all bits in B. Meanwhile, each bit in the between is a Boolean function in which the degree increases as the bit position gets closer to the MSB.

In order to know whether a particular selected monomial $t_I$ is a maxterm, this study applies the same BLR test as used in [9][7] with 300 random vectors. The framework of the side-channel cube attack on the Simeck32/64 block cipher is designed based on HW-SCCA in [15]. The framework has been designed using the Hamming-weight leakage model. There are two phases of attack namely, pre-processing phase and the online phase of attack. Firstly, in the pre-processing phase, random vectors (consists of key and plaintext) are input into the Simeck cryptographic algorithm. Then, the cipher is run and the internal states after each round have been archived for the determination of Hamming-weight. In this side-channel model, it is recommended that the most ideal Hamming-weight leakage is in the round with which the cipher begins achieving complete diffusion. This enables the adversary to find most of the key variables from low degree master polynomials in early rounds of the encryption process.

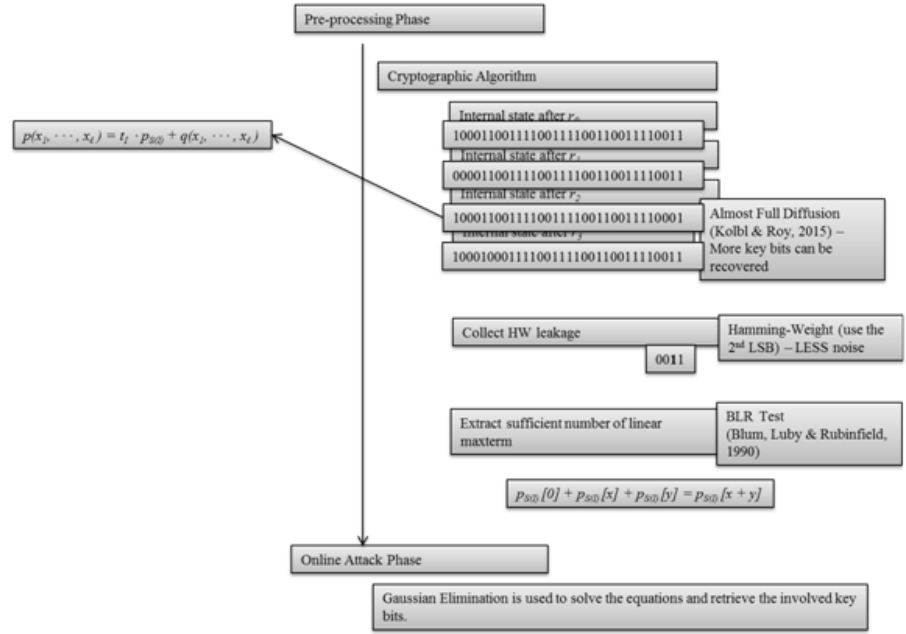

Figure 4. The proposed side-channel cube attack using Hamming-weight leakage model

As in Definition 1, in this study, the value of the Hamming-weight of the internal state after 4 rounds is represented in 8-bit binary form by considering all bits starting from LSB position towards the MSB position. Once the adversary has been able to collect Hamming-weight, in order to extract linear maxterm equation, $2^{nd}$ LSB of Hamming-weight has to be determined. In cube attack, to check whether a particularly selected monomial $t_I$ is a maxterm, this study applies the same BLR test as used in [7]. Finally, in the online phase, the adversary needs to solve all maxterm equations in order to recover secret keys. In the designed framework, Gaussian Elimination is recommended to be used.

In our study, the framework (see Figure 4) has been executed using simulation on several laptops with a running C program. Each of the laptops has been installed with a running C program with respective cube size. The recovered maxterm equations are save into text files (.txt) which then analyzed with Gaussian Elimination.

## 7. FINDINGS OF THE SIDE-CHANNEL CUBE ATTACK ON SIMECK32/64

After running the preprocessing phase of the attack for several weeks, thousands of maxterm equations using different cubes sizes have been collected, where most of them were found to be redundant and linearly dependent equations. To filter the equations and obtain only linearly independent equations among them, MATLAB and Gaussian Elimination are used. The elimination gives us only 32 linearly independent equations over 32 key variables. Table 3 shows the indexes of variables in the maxterms and the corresponding linearly independent equations that have been obtained. In this study, the indexes for both the plaintext and the key variables start from index 0, namely the MSB, until 63 (LSB). The total time complexity to find the correct 64-bit key reduces to $2^{35}$ compared to the $2^{64}$ for an exhaustive key search attack (brute force).

Table 3. Results of side-channel cube attack on Simeck32/64.

| Cube Indexes | Maxterm Equation |
|---|---|
| {0, 1, 2, 3, 4, 61} | $k_{11}$ |
| {0, 1, 2, 3, 6, 20} | $k_{12}$ |
| {0, 1, 2, 3, 15, 60} | $k_{10}$ |
| {0, 1, 2, 3, 57, 58} | $k_{15}$ |
| {0, 1, 2, 4, 21, 59} | $k_9$ |
| {0, 1, 2, 4, 58, 61} | $k_7$ |
| {0, 1, 2, 7, 19, 58} | $k_{10} + k_{14} + k_{13}$ |
| {0, 1, 2, 7, 56, 58} | $k_{10} + k_{13} + k_{14} + k_{12} + k_8$ |
| {0, 1, 2, 9, 26, 59} | $k_4$ |
| {0, 1, 2, 10, 27, 53} | $k_{13}$ |
| {0, 1, 2, 10, 27, 58} | $k_1$ |
| {0, 1, 2, 11, 14, 59} | $k_6$ |
| {0, 1, 2, 13, 19, 58} | $k_{14} + k_{14} + k_0 + k_8 + k_7 + k_3$ |
| {0, 1, 2, 13, 30, 58} | $k_{14} + k_0 + k_8$ |
| {0, 1, 3, 4, 20, 26} | $k_{14}$ |
| {0, 1, 3, 8, 20, 53} | $k_{14} + k_{13} + k_{12}$ |
| {0, 1, 3, 8, 25, 58} | $k_{10} + k_{14} + k_{13} + k_{14} + k_{13} + k_5$ |
| {0, 1, 3, 9, 17, 60} | $k_{14} + k_8$ |
| {0, 1, 3, 11, 58, 63} | $k_{12} + k_7 + k_6$ |
| {0, 1, 3, 13, 55, 58} | $k_2$ |
| {0, 1, 3, 14, 17, 59} | $k_{12} + k_{10} + k_0$ |
| {0, 1, 3, 14, 20, 58} | $k_{13} + k_{10} + k_{10} + k_9 + k_4 + k_4$ |
| {0, 1, 4, 12, 18, 56} | $k_{12} + k_7 + k_4 + k_7 + k_5 + k_2$ |
| {0, 1, 4, 14, 56, 58} | $k_3$ |
| {0, 1, 5, 17, 22, 60} | $k_7 + k_{21} + k_{12} + k_{11} + k_{10} + k_8$ |
| {0, 2, 3, 5, 19, 61} | $k_{17} + k_{12} + k_{11}$ |
| {0, 2, 3, 11, 19, 61} | $k_{12} + k_{14} + k_7 + k_6 + k_3 + k_1$ |
| {0, 2, 7, 29, 56, 58} | $k_1$ |
| {0, 4, 15, 16, 21, 59} | $k_{16} + k_{10} + k_{11} + k_{10} + k_9 + k_3$ |
| {1, 2, 3, 11, 28, 53} | $k_4 + k_0$ |
| {1, 2, 4, 9, 26, 59} | $k_{11} + k_{13} + k_{13} + k_{14} + k_{10} + k_0$ |
| {0, 1, 2, 4, 11, 15, 41, 46} | $k_{10} + k_{10} + k_{10} + k_2 + k_1$ |

As shown in Table 3, after four rounds (as shown in Figure 5(a)) of encryption, the maxterms start to appear within $t_i s$ of size 6; there are 31 maxterms of size 6 and 1 maxterm of size 8. Based on Figure 5(b), most of the sub keys are found within the cube size of 6. Hence, the total number of the chosen plaintexts for the online phase of the cube attack is $(31 \times 2^6) + (1 \times 2^8) \approx 2^{11.2855}$ by considering 32 linearly independent equations over the 32 key variables.

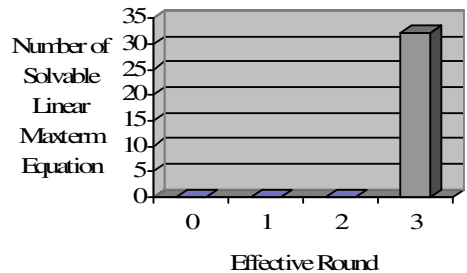

(a) Total of solvable maxterm equation based on effective round

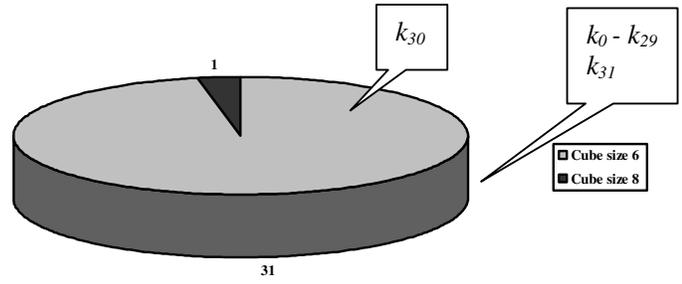

(b) Total of solvable maxterm equation and sub key bits after 4 rounds
encryption based on cube size

Figure 5. The appearance of sub key bits in solvable linear maxterm equation and in Simeck32/64

As the result, by using Hamming-weight leakage assumption in our side-channel cube attack on Simeck32/64, there are more sub key bits have been recovered compared to other method of attack in side-channel model (full round attack). The conventional method is the adversary has to recover the exact target bit in the internal state. Using Hamming-weight, the adversary has to calculate the Hamming-weight of the internal state only (more relax). Our findings show that the security evaluation on Simeck32/64 has been improved in terms of the time and data complexity. The analysis of side-channel attack on a block cipher is important for the considerations in the cryptosystem implementation. So, the implementer of the cryptosystem should prevent information from a leakage. Table 4 shows the comparison of results on Simeck32/64 in side-channel attack model. Other results are shown in Table 1.

Table 4. Comparison of results on Simeck32/64 in side-channel attack model.

| Method of Attack | Time Complexity |
|---|---|
| Our Method (Hamming-weight leakage) | $2^{35}$ |
| Single-bit fault | $2^{48}$ |
| Random-byte fault | $2^{48}$ |

## 8. CONCLUSIONS

This paper discusses the security of the Simeck32/64 block cipher against cube attacks with Hamming-weight leakage assumption in the side-channel attack model. The analysis shows that the attack can recover half of the 64-bit key of the cipher, by considering a Hamming-weight leakage bit (second LSB bit) from the internal state after the 4 rounds, with $2^{35}$ computations. The attack has been able to find 32 linearly independent equations over 32 key variables. However, there are some nonlinear equations of low degree (of degree 2) found during the analysis which may further reduce the complexity of the attack. To be implemented in IoT applications, further analysis is required. In addition to this study, further research and investigation on the other two variants of the Simeck family of block ciphers are strongly recommended.

### ACKNOWLEDGEMENTS

This work was supported by Universiti Teknologi MARA (UiTM) Malaysia under SLAB cholarship and Fundamental Research Grant Scheme of UTEM FRGS/1/2015/ICT05/FTMK/02/F00293 funded by Ministry of Higher Education, Malaysia.

## AUTHORS

**Alya Geogiana Buja** is a Ph.D. student at the Faculty of Information Technology and Communication, Universiti Teknikal Malaysia Melaka, Malaysia. Her research interests include information and network security. She involves actively in giving seminar and talk about information and security.

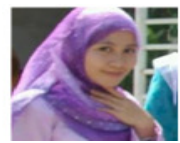

**Shekh Faisal Abdul-Latip** is a Senior Lecturer at the Faculty of Information Technology and Communication, Universiti Teknikal Malaysia Melaka, Malaysia. He received PhD degree in 2012 from the University of Wollongong, Australia, in the field of Symmetric-key Cryptography. Currently he is an executive committee member of Malaysian Society for Cryptology Research (MSCR) - a non-profit organization that promotes new ideas and activities in cryptology related areas in Malaysia. His research focuses on Cryptology (i.e. designing and breaking secret codes) and Information Security.

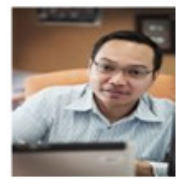

**Rabiah Ahmad** is a Professor at the Faculty of Information Technology and Communication, Universiti Teknikal Malaysia Melaka, Malaysia. She received her PhD in Information Studies (health informatics) from the University of Sheffield, UK, and M.Sc. (information security) from the Royal Holloway University of London, UK. Her research interests include healthcare system security and information security architecture. She has delivered papers at various health informatics and information security conferences at national as well as international levels. She has also published papers in accredited national/international journals. Besides that, she also serves as a reviewer for various conferences and journals.

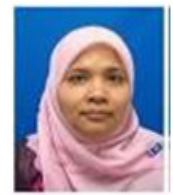